\documentclass[twocolumn,floatfix,prl,showpacs,superscriptaddress]{revtex4}
\usepackage{graphicx}

\usepackage{amsmath}
\usepackage{hyperref}
\graphicspath{{Graphs/}}

\begin{document}

\title{Evidence of one-dimensional magnetic heat transport in the triangular-lattice antiferromagnet Cs$_2$CuCl$_4$}
\author{E.~Schulze}
\affiliation{Dresden High Magnetic Field Laboratory (HLD-EMFL) and W\"urzburg-Dresden Cluster of Excellence ct.qmat, Helmholtz-Zentrum Dresden-Rossendorf, 01328 Dresden, Germany}
\affiliation{Institute of Solid State and Materials Physics, TU Dresden, 01062 Dresden, Germany}
\author{S.~Arsenijevic}
\affiliation{Dresden High Magnetic Field Laboratory (HLD-EMFL) and W\"urzburg-Dresden Cluster of Excellence ct.qmat, Helmholtz-Zentrum Dresden-Rossendorf, 01328 Dresden, Germany}
\author{L.~Opherden}
\affiliation{Dresden High Magnetic Field Laboratory (HLD-EMFL) and W\"urzburg-Dresden Cluster of Excellence ct.qmat, Helmholtz-Zentrum Dresden-Rossendorf, 01328 Dresden, Germany}
\affiliation{Institute of Solid State and Materials Physics, TU Dresden, 01062 Dresden, Germany}
\author{A.~N.~Ponomaryov}
\affiliation{Dresden High Magnetic Field Laboratory (HLD-EMFL) and W\"urzburg-Dresden Cluster of Excellence ct.qmat, Helmholtz-Zentrum Dresden-Rossendorf, 01328 Dresden, Germany}
\author{J.~Wosnitza}
\affiliation{Dresden High Magnetic Field Laboratory (HLD-EMFL) and W\"urzburg-Dresden Cluster of Excellence ct.qmat, Helmholtz-Zentrum Dresden-Rossendorf, 01328 Dresden, Germany}
\affiliation{Institute of Solid State and Materials Physics, TU Dresden, 01062 Dresden, Germany}
\author{T.~Ono}
\affiliation{Department of Physical Science, Osaka Prefecture University, Osaka 599-8531, Japan}
\author{H.~Tanaka}
\affiliation{Department of Physics, Tokyo Institute of Technology, Tokyo 152-8551, Japan}
\author{S.~A.~Zvyagin}
\affiliation{Dresden High Magnetic Field Laboratory (HLD-EMFL) and W\"urzburg-Dresden Cluster of Excellence ct.qmat, Helmholtz-Zentrum Dresden-Rossendorf, 01328 Dresden, Germany}

\date{\today}

\begin{abstract}

We report on low-temperature heat-transport properties of the spin-1/2 triangular-lattice antiferromagnet Cs$_2$CuCl$_4$.
Broad maxima in the thermal conductivity along the three principal axes,  observed at about 5 K,  are interpreted  in terms of
the Debye model, including the phonon Umklapp scattering. For thermal transport along the $b$ axis, we observed 
a pronounced field-dependent anomaly,  close to the transition into the three-dimensional long-range-ordered  state.  No such anomalies
were found for the transport along the $a$ and $c$ directions.
We argue that this  anisotropic  behavior is  related to an additional  heat-transport channel through magnetic excitations, that
can best propagate along the direction of the largest exchange interaction. Besides, peculiarities of the heat transport of Cs$_2$CuCl$_4$
in magnetic fields up to the saturation field and above are discussed.
\end{abstract}
\pacs{75.40.Gb, 76.30.-v, 75.10.Jm}
\maketitle

Frustrated magnets are known to be  an excellent playground to test  fundamental concepts of condensed matter physics  and quantum mechanics \cite{Ramirez, Balents, Lac, Andrei, Starykh_Rev, Wosnitza}.
Spin-1/2 Heisenberg antiferromagnets (AFs)  on triangular lattices have attracted particular  attention,  representing   an important class
of low-dimensional (low-D)  frustrated magnets and  allowing one  to probe  effects of the geometrical frustration,  magnetic order,
and  quantum fluctuations in strongly correlated spin systems. Particularly, this interest was stimulated by the idea of the ``resonating valence bond'' (RVB)
ground state for an AF system of spins on a triangular layer lattice \cite{Anderson}. This quantum-disordered ground state  was proposed to
be a 2D fluid of resonating spin-singlet pairs, with the elementary excitation spectrum formed by fractionalized mobile quasiparticles, spinons.
Since that time, searching for experimental realizations
of the 2D quantum spin liquid  appears to be one of the central topics  in quantum physics.

\begin{figure}
\centering
\includegraphics[width=0.4\textwidth]{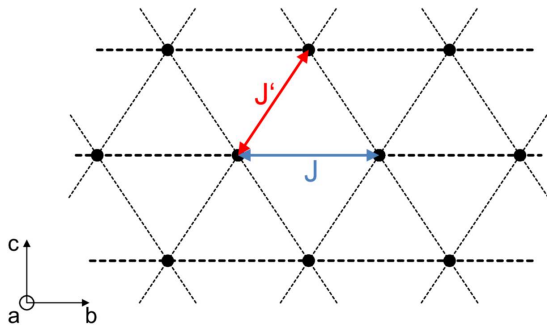}
\caption{Schematic view of exchange paths in Cs$_2$CuCl$_4$ with the exchange coupling  $J$ along the $b$  axis (chain direction) and $J'$ along the
zigzag bonds in the $bc$ plane.}
\label{structure}
\end{figure}

Among others,  the triangular-lattice AF Cs$_2$CuCl$_4$ has drawn  a particular  great deal of attention. Magnetic Cu$^{2+}$ ions in Cs$_2$CuCl$_4$
form a quasi-2D lattice with the exchange coupling $J$ along the $b$ axis (regarded as the chain direction) and $J'$ along the zigzag
bonds in the $bc$ plane (Fig. \ref{structure}). Its  spin Hamiltonian reads

\begin{align}
\mathcal{H}_{\text{2d}}=\sum_i J\vec{s}_i\cdot\vec{s}_{i+1}+\sum_k J'\vec{s}_k\cdot\vec{s}_{k+1} +\mathcal{H}_{\delta},
\end{align}
where $\vec s_i$ are the spins along and $\vec s_k$ between the chains (along the zigzag bonds), and $\mathcal{H}_{\delta}$ represents various possible,
usually small,   contributions (such as interlayer and the Dzyaloshinskii-Moriya (DM) interactions). High-field electron spin
resonance (ESR) measurements \cite{Zvyagin_J} revealed $J/k_B=4.7$ K and $J'/k_B=1.4$ K ($k_B$ is the Boltzman
constant), yielding the ratio $J'/J\approx 0.3$, which is in good agreement with parameters estimated from neutron-scattering experiments
\cite{Coldea3}.  The interlayer coupling   in Cs$_2$CuCl$_4$  appears to be smaller than $J$ and $J'$ by more than one order of magnitude, $J''=0.13$ K
\cite{Coldea3}.   At $T_N=0.62$ K,  Cs$_2$CuCl$_4$ undergoes a phase transition into a cycloidal  long-range-ordered
state  with an incommensurate wave vector $q$=(0,0.472,0) \cite{Coldea0}. Saturation fields of Cs$_2$CuCl$_4$ ($\mu_0
H_{sat} = 8.44$, 8.89, and 8 T, along the $a$, $b$, and $c$ axes, respectively \cite{Tokiwa}) can be reached using standard
superconducting magnets, allowing one to experimentally investigate the phase diagram in detail \cite{Coldea_2D_SL,Tokiwa,Hannahs}. 
 The observation of a number of subsequent low-temperature field-induced transitions have triggered intensive theoretical studies 
(\cite{Starykh_Rev} and references herein), revealing the important role of $\mathcal{H}_{\delta}$ (Eq. 1). 
 Very recently, several new field-induced transitions were observed in Cs$_2$CuCl$_4$,
emerging under applied hydrostatic pressure \cite{Zvyagin_pr}.

Inelastic neutron-scattering experiments on Cs$_2$CuCl$_4$ revealed the presence of a highly dispersive continuum of excited states
\cite{Coldea_2D_SL,Coldea_Cont}. These states have been initially identified as 2D RVB states, as suggested to occur in the AF system of spins on a triangular layer lattice 
\cite{Anderson}.  However, later on, the data have been re-interpreted in the framework of the quasi-1D  Tomonaga-Luttinger spin-liquid scenario 
\cite{Kohno}, with spinons (and  their interchain  bound-state excitations, triplons) as elementary magnetic excitations. 
 A key  condition here is the presence of the geometrical frustration, making spin chains  well-isolated from each other    
(this is in contrast to the 2D Majorana spin-liquid scenario \cite{Herf}, with magnetic excitations, however,   coherently propagating  
along the direction of the strongest exchange interaction). The quasi-1D  Tomonaga-Luttinger spin-liquid scenario perfectly describes the overall picture of magnetic excitations in  Cs$_2$CuCl$_4$, 
including 
their behavior in magnetic fields \cite{Kohno2}.  ESR studies have  supported   the proposed  model, with the uniform DM  interaction opening an energy gap ($\sim  0.7$ K)  at the $\Gamma$ 
point in the quantum-disordered  state \cite{Povarov}.

Here, we report on heat-transport studies of Cs$_2$CuCl$_4$, with the main goal to probe the anisotropy of the thermal conductivity on 
the temperature scale $T\sim J/k_{B}$,  where effects of the dimensionality of  magnetic correlations  should become relevant (\cite{Hess_rev} and references herein).

Single crystals of Cs$_2$CuCl$_4$ were grown by the slow evaporation of aqueous
solution of CsCl and CuCl$_2$ in the mole ratio 2 : 1. The room-temperature crystallographic  parameters ($a= 9.769$ \AA, $b= 7.607$ \AA,
~and $c=12.381$ ~\AA; ~space group $Pnma$)  were confirmed by x-ray diffraction and are  in good agreement with that   reported previously
\cite{Bailleul}.

Thermal-conductivity  measurements were performed in a $^3$He-cryostat, in magnetic fields up to 14 T. Samples were prepared  in a rod-like shape (with a thickness of about 0.5 mm 
and a length of about  5 mm) to allow  for a sufficiently high temperature gradient.
The temperature difference was produced
by a heater attached at one end of the sample and measured
by a pair of matched RuO$_2$ thermometers (as thermometers we used standard thick-film RuO$_2$-based SMD resistors);  a four-point technique was employed to measure the temperature
gradient along the sample. To reduce the statistical error, one point in the thermal conductivity (at  fixed  temperature and magnetic field) was averaged over 20 measurement cycles. 
In the experiments, the magnetic field was applied along  the direction of the heat transport. The temperature in the entire field range  was measured with accuracy 
better than $\pm 5\%$.

\begin{figure}
\centering
\includegraphics[width=0.43\textwidth]{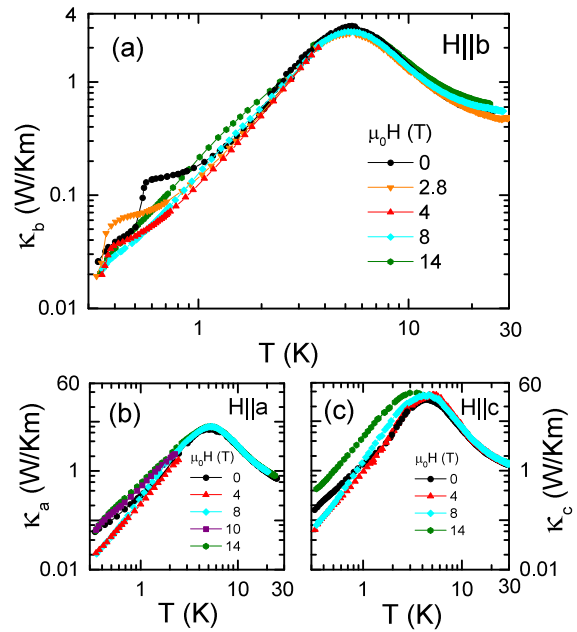}
\caption{Temperature dependence of the thermal conductivity  $\kappa_b$ (a), $\kappa_a$ (b), and $\kappa_c$ (c)    along the $b$, $a$, and $c$  axes, respectively,  in  selected magnetic fields. 
Lines are guides to the eye.}
\label{tempsweeps_b}
\end{figure}

The thermal-conductivity  experiments on Cs$_2$CuCl$_4$ revealed a broad maximum at about 5 K for all three directions
(Fig. \ref{tempsweeps_b}).  Such a behavior is  a text-book example of the phonon-dominated thermal
conductivity, where the low-temperature conductivity is determined by phonon scattering on  the crystal boundaries and   structural
imperfections, while the high-temperature  mean-free path is limited  by  the Umklapp  scattering   \cite{Kittel}. As one can see,
 absolute values of the thermal conductivity (in particular at the maximum positions) are different for different directions: it is
maximal for the direction along the $c$ axis  and few times smaller  for the $a$ and $b$ directions. This difference  can be tentatively explained
by the anisotropy of sound velocity  and phonon-magnon scattering  in Cs$_2$CuCl$_4$ at low temperatures.

With  decreasing temperature we observed  that the heat-transport behavior along the $b$ direction becomes 
significantly different from that along the $a$ and $c$ directions.  Along the $b$ axis, before entering the  3D ordered phase,
the thermal conductivity changes the slope (reflecting its pronounced  enhancement)  and  suddenly drops, once the material undergoes the 3D ordering. The anomaly position
depends on the applied magnetic field, shifting to lower temperatures with increasing field (which is  consistent with the temperature-field 
phase diagram obtained by Tokiwa $et ~al.$ \cite{Tokiwa}). 
Based on these observations, we suggest  that the  anomaly  of the thermal conductivity,  
revealed by us   in the vicinity of $T_N$   specifically 
along the $b$ direction,   has a magnetic  nature and is determined by magnetic excitations propagating along the 
direction of the largest exchange interaction (chain direction). For the Tomonaga-Luttinger spin liquid, proposed for  Cs$_2$CuCl$_4$ \cite{Kohno},  these excitations are spinons.  
Similar  anisotropic behavior of the thermal transport was observed  in quasi-1D chain materials  
Sr$_2$CuO$_3$ \cite{Sol_1,Sol_2}, SrCuO$_2$ \cite{Sol_2},  Cu(C$_4$H$_4$N$_2$)(NO$_3$)$_2$
\cite{Sol_3}, and CaCu$_2$O$_3$ \cite{Hess2}.
Noticeably, apart from the shift, the applied magnetic field suppresses  the anomaly, making it almost undetectable at 8 T. 
Such a behavior is consistent with  the field-induced  crossover from the   quantum  to  a
classically-favored state, where  quantum fluctuations are significantly suppressed by magnetic field  \cite{Starykh_Jin}.   The sudden drop of the thermal conductivity $\kappa_b$ at $T_N$ can 
be explained by a collapse 
of the 1D spinon-heat transport when entering the 3D AF ordered state, 
accompanied by the opening of an   energy gap in the low-temperature excitation spectrum. Such a gap ($\sim 1.3$ K) was observed in 
Cs$_2$CuCl$_4$ by means of ESR \cite{Smirnov}.

\begin{figure}
\centering
\includegraphics[width=0.47\textwidth]{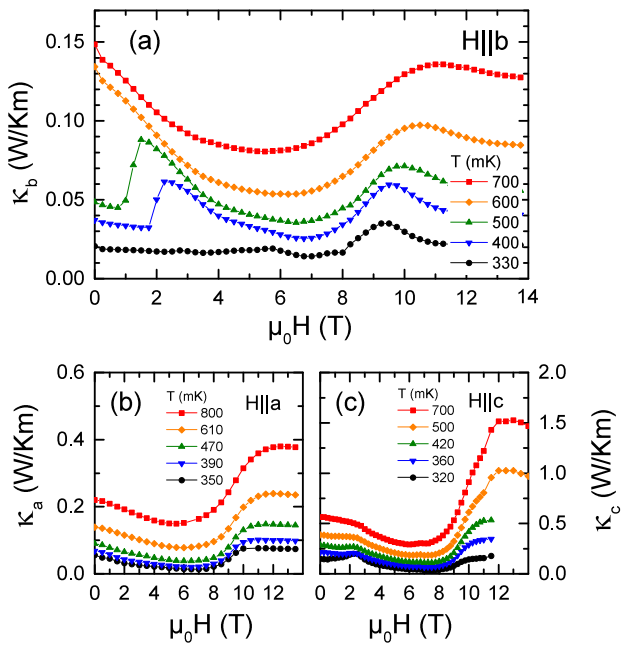}
\caption{Field dependence of the thermal conductivity  $\kappa_b$ (a), $\kappa_a$ (b), and $\kappa_c$ (c) at selected temperatures.
The magnetic field was applied along the heat-transport directions.
 Lines are guides to the eye.}
\label{fieldsweeps_abc}
\end{figure}

Field measurements  revealed a  non-monotonic behavior of the thermal conductivity for all three directions of the applied magnetic fields 
(Fig. \ref{fieldsweeps_abc}). For the heat transport along the $b$ and $c$ axes,  we observe  changes
in the heat transport behavior,  which occur in magnetic fields about  2 T. These anomalies correspond to field-induced transitions 
from the 3D ordered   to a disordered  state, as reported previously \cite{Coldea_2D_SL,Tokiwa,Hannahs}. Noticeable  low-field jumps of 
$\kappa_b$ at 
about  0.5 and 0.4 K (Fig. \ref{fieldsweeps_abc}a)
 provide  additional evidences of the 1D
nature of magnetic excitations   in this material, significantly contributing to the heat  transport just above $T_N$.

\begin{figure}
\centering
\includegraphics[width=0.45\textwidth]{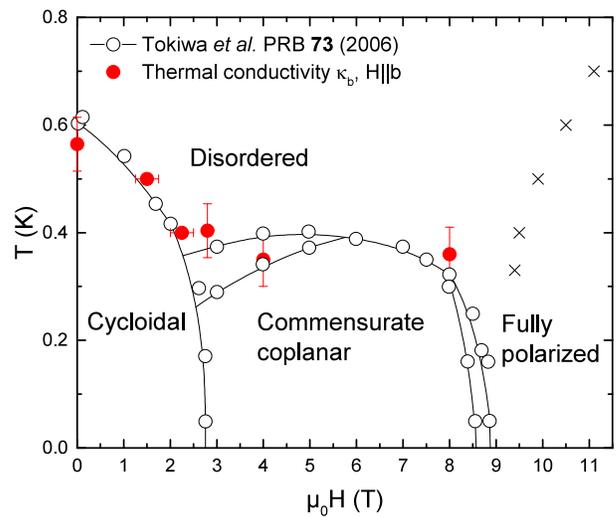}
\caption{Phase diagram of Cs$_2$CuCl$_4$ for magnetic field applied along the $b$ direction. Data from our thermal-transport measurements
are shown by red solid circles, while specific-heat and susceptibility data from \cite{Tokiwa}  are denoted by open circles. The crosses correspond to the
thermal conductivity  maxima in the fully spin-polarized phase above $H_{sat}$ (Fig. \ref{fieldsweeps_abc}). Lines are guides to the eye. }
\label{phasediagram}
\end{figure}

Another peculiar finding  is the pronounced dip in the filed dependences of thermal conductivity at about 6-7 T, observed for all three directions 
(Fig. \ref{fieldsweeps_abc}). This observation is consistent with the low-temperature ultrasound  properties of Cs$_2$CuCl$_4$ \cite{Sytcheva,Streib,Tsyp}, suggesting the 
exchange-striction mechanism as a possible reason of such a behavior.  A pronounced increase of 
the thermal conductivity above  $H_{sat}$ can be understood taking into account the decreased phonon-magnon  scattering in the
fully spin-polarized phase (the positions of  corresponding maxima in thermal conductivity along the $b$ direction 
 are shown in Fig. \ref{phasediagram} by crosses).  The phase diagram obtained by us, together with results of previously reported  specific-heat and 
magnetic-susceptibility studies \cite{Tokiwa}
for $H\|b$, is shown in  Fig. \ref{phasediagram}; excellent agreement between the two data sets is found.

\begin{figure}
\centering
\includegraphics[width=0.47\textwidth]{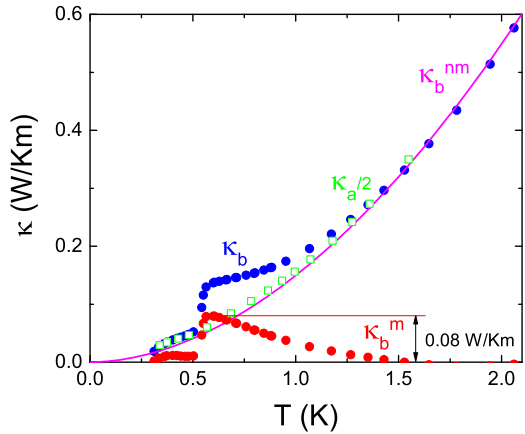}
\caption{Temperature dependences of the thermal conductivity $\kappa_b$ along the $b$ axis  (rough experimental data; blue symbols), nonmagnetic  background $\kappa_b^{nm}$ as results of the power law fit 
(magenta line), magnetic contribution to the thermal conductivity $\kappa_b^m$ as result of the subtraction $\kappa_b^{nm}$ from $\kappa_b$ (red symbols), and rescaled thermal conductivity 
$\kappa_a/2$ along the $a$ direction (green symbols).}
\label{anomaly}
\end{figure}

To estimate the magnetic mean-free path $l_b^{m}$ along the $b$ direction, low-temperature  heat transport properties of Cs$_2$CuCl$_4$ were 
analyzed in the framework   of the spin-1/2 AF Heisenberg chain model  \cite{Hess2} with  spinons as  elementary  magnetic 
excitations:

\begin{align}
l_b^m=\frac{\kappa_b^m}{T}\cdot\frac{3\hbar}{N_s k_B^2\pi}.
\end{align}
Here,  $\kappa_b^{m}$ is the magnetic heat conductivity,  $\hbar$ is the reduced Planck constant, 
and $N_s=4/ac$ is the number of spins per unit area. To estimate the magnetic heat conductivity $\kappa_b^m$ the nonmagnetic background $\kappa_b^{nm}$ was subtracted from the experimental data $\kappa_b$.
To obtain  $\kappa_b^{nm}$, the zero-field thermal-conductivity data $\kappa_b$   below and above the anomaly were fitted with the power law  
$\kappa = 0.1496\cdot T^{1.875}$ (line $\kappa_b^{nm}$ in  Fig. \ref{anomaly}). For comparison, the rescaled thermal conductivity along the $a$ direction $\kappa_a/2$ (where no anomaly was observed) 
 is shown, revealing excellent agreement with the fit results. 
The subtraction results yield 0.08 W/Km for the magnetic part of the heat conductivity   at the maximum position.  
Based on this model,we obtained $l_{m}\approx 200$ ~\AA,   which corresponds to approximately 30 lattice spacings 
along the $b$ direction.

In conclusion, the thermal conductivity in Cs$_2$CuCl$_4$ was measured   at temperatures down to 300 mK in magnetic fields up to 14 T
along the three principal crystallographic directions.  The heat transport is found to be dominated by the phonon contribution. 
On the other hand, a pronounced
field-dependent anomaly of the thermal conductivity  was observed  along the $b$ axis,  when approaching the transition into the
3D ordered   state. The anomaly is attributed to the  1D heat transport through  magnetic excitations propagating in Cs$_2$CuCl$_4$  along the 
direction of the strongest exchange coupling. Our observations strongly support the  quasi-1D  spin-liquid scenario with spinons as elementary excitations, proposed for 
this frustrated antiferromagnet.  

\textit{Acknowledgements.} This work was supported by the Deutsche Forschungsgemeinschaft (DFG), through ZV 6/2-2, the excellence
cluster ct.qmat (EXC2147, project-id 39085490), and SFB 1143, as well as by the HLD at HZDR, member of the European Magnetic Field Laboratory (EMFL).
We acknowledge fruitful discussions with O.~Starykh, T.~Lorenz,  X. Zotos,  A.~ Chernyshev, A.~Zheludev,  S.~Zherlitsyn,  U.~H.~Acosta, and 
A.~Pidatella.

\end{document}